\documentclass[sn-mathphys-num]{sn-jnl}

\usepackage{amsmath,amsfonts}
\usepackage{algorithmic}
\usepackage{algorithm}
\usepackage{array}
\usepackage[caption=false,font=normalsize,labelfont=sf,textfont=sf]{subfig}
\usepackage{textcomp}
\usepackage{stfloats}
\usepackage{url}
\usepackage{verbatim}
\usepackage{graphicx}
\usepackage{pdfpages}
\usepackage{enumitem}
\usepackage{natbib}

\usepackage{amssymb}
\usepackage{amsthm}
\usepackage{booktabs}
\usepackage{threeparttable}
\usepackage{braket}

\usepackage{tikz}
\usetikzlibrary{arrows.meta, positioning, shapes.multipart}

\tikzset{
  prismaBox/.style={
    draw, rounded corners, align=center, minimum width=0.9\linewidth,
    inner sep=6pt, font=\normalsize
  },
  smallBox/.style={
    draw, rounded corners, align=center, minimum width=0.9\linewidth,
    inner sep=4pt, font=\normalsize
  },
  arrow/.style={-Latex, very thick}
}

\theoremstyle{thmstyleone}%
\theoremstyle{thmstyletwo}%

\theoremstyle{thmstylethree}%

\begin{document}

\title[Quantum Machine Unlearning: Foundations, Mechanisms, and Taxonomy]{Quantum Machine Unlearning: Foundations, Mechanisms, and Taxonomy}


\author*[1]{\fnm{Thanveer} \sur{Shaik}}\email{thanveer.shaik@unisq.edu.au}

\author[1]{\fnm{Xiaohui} \sur{Tao}}\email{xiaohui.tao@unisq.edu.au}
\equalcont{These authors contributed equally to this work.}


\author[2]{\fnm{Haoran} \sur{Xie}}\email{hrxie@ln.edu.hk}
\equalcont{These authors contributed equally to this work.}

\affil[1]{\orgdiv{School of Science, Engineering, and Digital Technologies}, \orgname{University of Southern Queensland}, \orgaddress{\country{Australia}}}
\affil[2]{\orgdiv{School of Data Science}, \orgname{Lingnan University}, \orgaddress{\country{Hong Kong}}}




\abstract{Quantum Machine Unlearning (QMU) has emerged as a foundational challenge at the intersection of quantum information theory, privacy-preserving computation, and trustworthy artificial intelligence.  
This paper advances QMU by establishing a formal framework that unifies physical constraints, algorithmic mechanisms, and ethical governance within a verifiable paradigm.  
We define forgetting as a \emph{contraction of distinguishability} between pre- and post-unlearning models under completely positive trace-preserving (CPTP) dynamics, grounding data removal in the physics of quantum irreversibility.  
Building on this foundation, we present a five-axis taxonomy spanning scope, guarantees, mechanisms, system context, and hardware realization, linking theoretical constructs to implementable strategies.  
Within this structure, we incorporate influence- and quantum Fisher information–weighted updates, parameter reinitialization, and kernel alignment as practical mechanisms compatible with noisy intermediate-scale quantum (NISQ) devices.  
The framework extends naturally to federated and privacy-aware settings through quantum differential privacy, homomorphic encryption, and verifiable delegation, enabling scalable and auditable deletion across distributed quantum systems.  
Beyond technical design, we outline a forward-looking research roadmap emphasizing formal proofs of forgetting, scalable and secure architectures, post-unlearning interpretability, and ethically auditable governance.  
Together, these contributions elevate QMU from a conceptual notion to a rigorously defined and ethically aligned discipline, bridging physical feasibility, algorithmic verifiability, and societal accountability in the emerging era of quantum intelligence.}


\maketitle

\section{Introduction}
\label{sec:intro}

Quantum computing and machine intelligence are converging to redefine how data-driven systems learn, adapt, and comply with privacy laws. As quantum machine learning (QML) transitions from small-scale prototypes to distributed deployments, the notion of the \emph{right to erasure} acquires a new dimension.  
Under frameworks such as the GDPR, erasure must extend beyond model parameters to include quantum states, channels, and measurements~\cite{cao2015towards,bourtoule2021machine}. In this setting, unlearning is not a simple rollback, it is a physically valid transformation that decouples private signals from global quantum behaviour. Quantum mechanics offers natural privacy safeguards through no-cloning, measurement disturbance, and decoherence, yet these same principles make verification, audit, and decentralized enforcement challenging at scale.

Federated learning (FL) decentralizes training but still exposes gradients and aggregates to inversion risks. Quantum Federated Learning (QFL) pushes privacy deeper into the computational stack by encoding updates as quantum states and coordinating learning through variational circuits~\cite{qiao2025transitioning}.  
When combined with quantum differential privacy (QDP), QFL can achieve rigorous confidentiality with minimal utility loss. Rofougaran~\textit{et al.}~\cite{rofougaran2024federateddp} demonstrate a QFL–QDP pipeline achieving $>98\%$ accuracy under $\epsilon<1.3$ on NISQ hardware, showing that privacy and performance can coexist. This privacy posture now extends to the transport layer. Quantum cloud infrastructures and 6G space–ground networks rely on quantum key distribution (QKD) and homomorphic encryption to guarantee that deleted states remain computationally and physically unrecoverable. Zeng~\textit{et al.} introduce quantum circuit random reconstruction homomorphic encryption (QCRRA), enabling encrypted circuit execution even on untrusted servers~\cite{zeng2026quantumcloud}. Similarly, hybrid QFL combined with QKD improves both anomaly detection and key generation rates, indicating that privacy-preserving protocols can enhance, not hinder, performance~\cite{mujlid2025quantumiot}.

Quantum deployments also expand the threat model.  
Adversaries may conduct membership inference, inversion, or poisoning attacks on parametrized quantum circuits, exploiting universal perturbations that transfer across architectures~\cite{qiu2023uap}.  
A defensible system must reason about privacy at the level of quantum channels.  
A natural analogue of $(\varepsilon,\delta)$-differential privacy requires that, for neighbouring datasets encoded as states and for any POVM, the resulting output distributions differ by at most $(\varepsilon,\delta)$ after the full CPTP pipeline.  
Composition across rounds and clients then follows from channel algebra.  
In practice, calibrated dephasing or depolarizing noise acts as privacy randomization, while error mitigation restores performance.  
In federated setups, cryptographic transport is indispensable.  
Measurement-device-independent QKD provides composable keys for authenticated encryption and Pauli one-time pads during aggregation, as illustrated in smart-grid benchmarks~\cite{ren2024qqfl}.  
Privacy must also persist through time.  
Quantum continual learning reduces catastrophic forgetting and constrains leakage from past tasks by allocating disjoint PQC subspaces or projecting gradients to avoid interference~\cite{xu2024dynamic,situ2023qcl}.  
Robustness-aware training, randomized encodings, and adversarial PQC optimization further limit attack transferability and restrict what can be inferred from exchanged statistics.

\textbf{Motivating scenarios.}  
Quantum machine unlearning (QMU) is motivated by diverse practical contexts.  
(i)~\emph{Healthcare withdrawal}: a patient revokes consent after a QFL oncology study, requiring the removal of that client’s contribution without retraining heterogeneous QPUs, while preserving model utility on retained cohorts.  
(ii)~\emph{Incident response}: a poisoning attack is detected in a quantum-enhanced intrusion detector; the system must excise tainted class components or client registers using physically valid channels and produce auditable evidence of deletion.  
(iii)~\emph{Cross-cloud portability}: workloads migrate between superconducting and trapped-ion backends, demanding device-agnostic proofs that residual entanglement cannot reveal client data.  
(iv)~\emph{Regulatory audit}: a national archive requests verifiable deletion of specific contributions to a quantum language model; the operator must supply DP accounting, forgetting curves, and channel-level certificates of contraction.  
Across these cases, unlearning means more than data deletion, it entails certified suppression of retrievable information under CPTP dynamics, consistent with federated orchestration and cryptographic transport.

\textbf{Review objectives.}  
This survey unifies the conceptual, algorithmic, and system-level foundations of QMU.  
We bridge classical unlearning principles with quantum information theory, propose a five-axis taxonomy that links scope, guarantees, mechanisms, system context, and hardware, and formalize forgetting as the contraction of distinguishability under CPTP maps.  
Our framework integrates privacy engineering techniques such as QDP, QKD-backed transport, and homomorphic execution with learning-theoretic and thermodynamic perspectives.  
The aim is to advance QMU from empirical prototypes toward verifiable, auditable, and scalable infrastructures.

\textbf{Contributions.}  
(1)~A \emph{quantum information formalization} of unlearning that replaces destructive erasure with CPTP-driven redistribution, anchored in the data-processing inequality and thermodynamic irreversibility.  
(2)~A \emph{taxonomy and evaluation framework} covering sample-, class-, and client-level scopes; empirical, certified, and DP guarantees; and mechanisms including reset, Fisher-weighted updates, and gradient reversal across QML and QFU deployments.  
(3)~\emph{Federated QMU protocols} that combine gradient hiding, differential privacy accounting, and cryptographic transport for secure orchestration~\cite{li2024privacy,bhatia2024irps,mashetty2025pqfl,ballester2025qflsurvey}.  
(4)~\emph{Threat-informed design principles} addressing transfer attacks and catastrophic forgetting through continual-learning and robustness-aware PQC training~\cite{qiu2023uap,xu2024dynamic,situ2023qcl}.  
(5)~\emph{Empirical evidence} that privacy and utility can coexist in practice, including hybrid recurrent and time-series models where variational quantum embeddings improve forecasting with compact architectures~\cite{yu2023qlstm}.

\textbf{Organization.}  
Section~\ref{sec:foundations} establishes the quantum-information foundations of forgetting as a CPTP contraction with operational metrics.  
Section~\ref{sec:taxonomy} introduces the unified taxonomy linking theoretical dimensions to system and hardware contexts.  
Section~\ref{sec:from-learning-to-unlearning} explores algorithmic pathways, from QFI-weighted influence updates to partial retraining and certification pipelines.  
Section~\ref{sec:eval} outlines evaluation metrics and benchmark datasets for reproducible, hardware-grounded studies.  
Section~\ref{sec:open-challenges} synthesizes the open challenges and research agenda spanning formal proofs, scalable architectures, interpretability, and ethical governance.  
We conclude with recommendations for building verifiable and privacy-preserving quantum learning ecosystems.

\section{Quantum Information Foundations of Unlearning}
\label{sec:foundations}

\subsection{From Classical to Quantum Unlearning}

Machine unlearning (MU) aims to remove the influence of selected training data without retraining the entire model.  
The idea aligns with legal and ethical frameworks such as the European Union’s \emph{Right to Erasure} in the GDPR~\cite{cao2015towards,bourtoule2021machine}, which demand that data and its effects be verifiably erased.  
Early approaches used practical engineering strategies, including checkpoint rollback and incremental updates.  
Influence functions introduced a more principled approach by estimating each sample’s contribution to the optimum through $-\mathbf{H}^{-1}\nabla_\theta L(\theta^*,x_i)$, with $\mathbf{H}$ as the empirical Hessian~\cite{koh2017understanding}.  
The SISA framework (\emph{Sharded, Isolated, Sliced Aggregation}) further improved auditability and deletion latency through modular retraining~\cite{bourtoule2021machine}.  

Three common goals unify this body of work.  
\emph{Efficiency} demands minimal computational cost and wall-clock time.  
\emph{Completeness} ensures both data and its downstream influence are removed.  
\emph{Verifiability} requires clear, measurable evidence of forgetting.  
Certified low-rank corrections enable exact updates in convex models~\cite{guo2020certified}, while approximate regularization and Bayesian re-weighting extend unlearning to deep, non-convex models by transitioning from $p(\theta|D)$ to $p(\theta|D\setminus D_r)$~\cite{nguyen2022bayesian}.  

Federated unlearning applies similar ideas to distributed learning.  
Each client locally compensates for removed data by adjusting its gradient contribution, e.g., $\theta'=\theta-\alpha\nabla_\theta L_i$, often using stored or compressed updates~\cite{liu2021federated}.  
Hybrid quantum–classical systems now adopt comparable coordination strategies.  
Quantum feature extractors integrate with classical aggregators~\cite{bhatia2023federated}, ring topologies support bandwidth-efficient communication~\cite{wang2024quantum}, and dynamic aggregation improves robustness under non-IID healthcare data~\cite{qu2025daqfl}.  
Information-theoretically, ideal unlearning reduces mutual information between the removed data and the model, $I(D_r;M)$, while retaining $I(D_s;M)$ for the preserved set~\cite{ginart2019making}.  
Metrics such as forgetting accuracy, retention, and retraining cost quantify progress toward this goal.  

These classical advances define a conceptual baseline.  
However, when computation occurs on quantum systems, forgetting must respect the laws of quantum information.  
This transition, from statistical optimization to physical constraint, marks the foundation of quantum machine unlearning (QMU).

\subsection{Quantum Learning Dynamics}

Quantum machine learning (QML) trains parameterized quantum circuits (PQCs) that encode data into quantum states and iteratively update parameters through measurement and feedback.  
Each PQC implements a unitary $U(\theta)$ acting on $\ket{0}^{\otimes n}$, and gradients are computed on hardware through the parameter-shift rule,
\[
\partial_{\theta_k}\mathbb{E}[O]
= \tfrac{1}{2}\big(\mathbb{E}[O]_{\theta_k+\frac{\pi}{2}}
- \mathbb{E}[O]_{\theta_k-\frac{\pi}{2}}\big),
\]
which allows exact differentiation under real hardware noise~\cite{schuld2019evaluating}.  

Model expressivity depends on data encoding and circuit depth.  
Re-uploading strategies expand hypothesis space by alternating data injection with trainable layers~\cite{fan2022compact}, while differentiable quantum programming provides compiler-level support for gradient computation across control structures~\cite{zhu2020pldi}.  
Training remains challenging due to \emph{barren plateaus}, where gradient variance decays exponentially with circuit depth~\cite{mcclean2018barren}.  
Geometry-aware optimization addresses this by using the quantum Fisher information (QFI) to rescale gradients,  
$\theta \leftarrow \theta - \eta F(\theta)^{-1}\nabla_\theta\mathcal{L}$,  
thereby improving convergence and shot efficiency on NISQ devices~\cite{stokes2020quantum}.  

Distributed learning frameworks extend these foundations.  
In quantum federated learning (QFL), multiple clients train local PQCs or hybrid models and share updates under privacy-preserving protocols such as gradient hiding, quantum differential privacy, or entanglement-based coordination~\cite{li2024privacy,bhatia2024irps,mashetty2025pqfl}.  
Comprehensive surveys document the architectural and privacy challenges that shape scalable QFL systems~\cite{ballester2025qflsurvey}.  
Collectively, exact gradient evaluation, QFI-guided optimization, and privacy-preserving coordination provide the algorithmic substrate on which QMU is built.  
They allow a quantum model to not only learn from data but also to forget it through controlled, measurable influence reduction.

Table~\ref{tab:classical-vs-quantum} contrasts classical and quantum MU along representation, mechanisms, feasibility limits, metrics, thermodynamic implications, and federated operation.

\subsection{Information Constraints: No-Cloning and No-Deletion}

Quantum information cannot be freely copied or erased.  
The \emph{no-cloning theorem} forbids a universal operation $\ket{\psi}\ket{0}\mapsto\ket{\psi}\ket{\psi}$~\cite{wootters1982noclone,dieks1982communication}, and the \emph{no-deletion theorem} prohibits $\ket{\psi}\ket{\psi}\mapsto\ket{\psi}\ket{0}$~\cite{pati2000nodelete}.  
These results align with \emph{no-broadcasting}~\cite{barnum1996nobroadcast} and \emph{no-hiding}~\cite{braunstein2007nohiding}.  
Therefore, perfect duplication or complete erasure is unphysical.  
Deletion in quantum systems can only occur as redistribution of information into an environment $E$:
\[
\ket{\psi}\ket{\psi}\ket{0}_E
\xrightarrow{U}
\ket{\psi}\ket{\mathrm{junk}(\psi)}_E.
\]
The environment now holds correlations that encode the “forgotten’’ content.  
This shift from removal to redistribution fundamentally redefines what forgetting means in the quantum setting.

\begin{table}[t]
\centering
\caption{Comparison between Classical and Quantum Machine Unlearning.}
\label{tab:classical-vs-quantum}
\begin{tabular}{p{0.27\columnwidth} p{0.32\columnwidth} p{0.32\columnwidth}}
\toprule
\textbf{Aspect} & \textbf{Classical Machine Unlearning (MU)} & \textbf{Quantum Machine Unlearning (QMU)} \\ 
\midrule
\textbf{Information representation} &
Parameters $\theta$ in Euclidean spaces. &
Density operators $\rho(\theta)$ with amplitudes and entanglement. \\
\textbf{Forgetting mechanism} &
Retraining, influence updates, low-rank corrections. &
CPTP channels $\mathcal{E}$ that decohere or redistribute information to $E$. \\
\textbf{Feasibility limits} &
No physical barrier to duplication/deletion. &
No-cloning and no-deletion; only redistribution/hiding is allowed. \\
\textbf{Mathematical formalism} &
$-\mathbf{H}^{-1}\nabla_\theta L(x_i)$; rollback in parameter space. &
$\rho'=\mathcal{E}(\rho)$; contractive evolution under quantum channels. \\
\textbf{Verification metric} &
Loss gaps; Euclidean/KL similarity; forgetting accuracy. &
Trace distance or fidelity to counterfactual $\rho^{\setminus D_r}$; data-processing inequality. \\
\textbf{Thermodynamic implication} &
Rarely modeled. &
Landauer cost: increased entropy and heat dissipation in $E$. \\
\textbf{Distributed/federated context} &
Subtract or reweight client gradients. &
Client-local maps $\mathcal{U}_c=(\mathcal{E}_c\!\otimes\!\mathrm{Id}_{\neg c})$ for partial disentanglement. \\
\bottomrule
\end{tabular}
\end{table}

Logical erasure also has energetic cost.  
Landauer’s principle states that each bit of erased information dissipates heat~\cite{landauer1961irreversibility,bennett2003notes}.  
In QMU, forgetting increases entropy in the environment rather than annihilating stored information.  
For any completely positive trace-preserving (CPTP) map $\mathcal{E}$ and states $\rho,\sigma$, the data-processing inequality,
\[
D(\rho\Vert\sigma)\ge D(\mathcal{E}(\rho)\Vert\mathcal{E}(\sigma)),
\]
ensures that distinguishability monotonically decreases under physical evolution~\cite{lindblad1975entropy,uhlmann1977relative}.  
Trace distance and fidelity exhibit the same contractive behavior, forming the mathematical foundation for certified quantum forgetting.  

\begin{figure}[t]
  \centering
  \includegraphics[width=0.95\columnwidth]{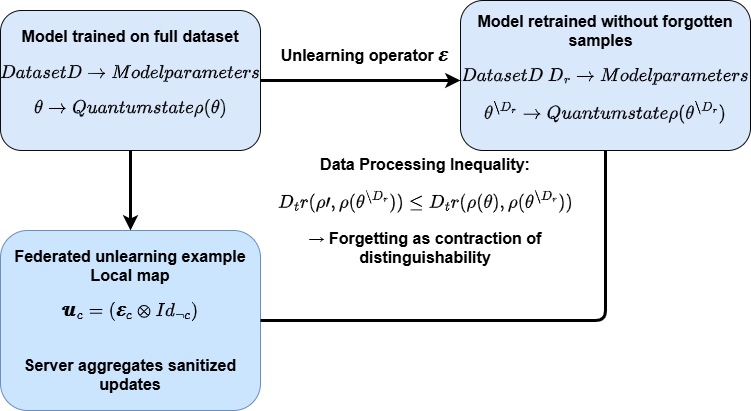}
  \caption{An unlearning channel $\mathcal{E}$ contracts the trace distance between models trained with and without the removed data, illustrating forgetting as redistribution under CPTP dynamics.}
  \label{fig:qmu-cptp}
\end{figure}

These physical limits redefine the purpose of unlearning.  
QMU cannot erase information; it instead makes the information operationally inaccessible.  
Practical pipelines combine three elements.  
First, CPTP channels provably contract trace distance or increase fidelity relative to the retrained counterfactual.  
Second, QFI-guided optimization reduces parameter sensitivity to individual samples.  
Third, federated and secure protocols prevent reconstruction through gradient hiding or differential privacy~\cite{joshi2023blind,shenoy2020measurement,chen2024robust,li2024privacy,ren2024tase}.  
Adversarial analyses confirm that measurable contraction, not ideal erasure, defines robust unlearning under realistic noise~\cite{kundu2025glsvlsi,ergu2025qpoison}.  

Fig.~\ref{fig:qmu-cptp} depicts this view: an unlearning operator $\mathcal{E}$ acts on the model state $\rho(\theta)$ and produces $\rho'=\mathcal{E}(\rho(\theta))$ that lies closer, in trace distance or equivalently higher in fidelity, to the counterfactual $\rho(\theta^{\setminus D_r})$ by data processing. In this sense, certified quantum forgetting is the controlled, physically consistent \emph{contraction} of the gap to the retrained state.

Together, these principles transform unlearning from an algorithmic adjustment to a physical process.  
They define the boundaries of what can be forgotten and provide the theoretical foundation for the next section, where we organize QMU methods through a unified taxonomy and evaluation framework.

\section{Taxonomy of Methods}
\label{sec:taxonomy}

\begin{figure}[t]
    \centering
    \includegraphics[width=\linewidth]{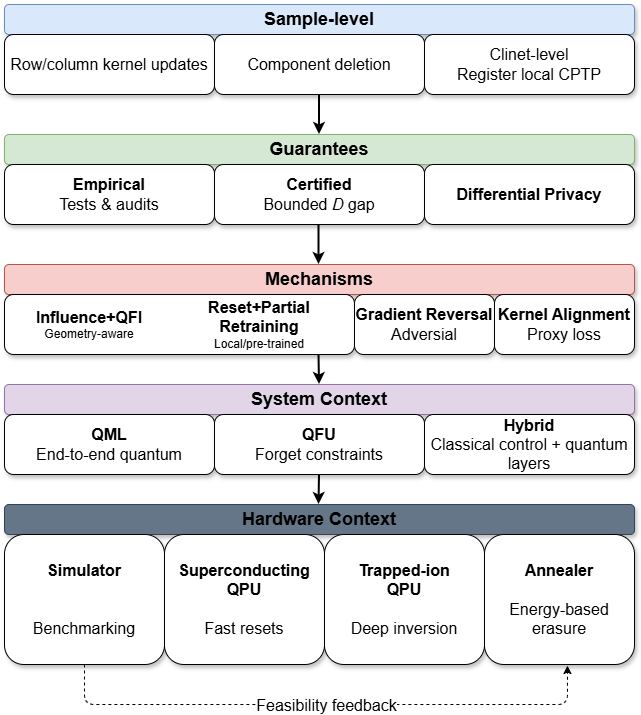}
    \caption{Unified taxonomy for quantum unlearning across five dimensions: scope, guarantees, mechanisms, system context, and hardware context.}
    \label{fig:taxonomy}
\end{figure}

Quantum machine unlearning (QMU) can be organized along five axes: \emph{scope}, \emph{guarantees}, \emph{mechanisms}, \emph{system context}, and \emph{hardware context}.  
These axes capture what we forget, how we demonstrate it, where we enact it, and under which physical constraints it remains feasible.  
Taken together, they link information-theoretic goals to implementable strategies and prepare the ground for the algorithmic pathways in Sec.~\ref{sec:from-learning-to-unlearning}. The taxonomy in Fig.~\ref{fig:taxonomy} joins theory and practice.  
Scope clarifies the target of deletion.  
Guarantees determine how we measure success.  
Mechanisms turn guarantees into edits.  
System and hardware contexts ensure those edits remain feasible and auditable on real devices.  

\subsection{Scope: Sample, Class, and Client Levels}
\label{subsec:scope}

Scope specifies the unit of forgetting.  
At the \emph{sample level}, the target set $\mathcal{S}=\{x_j\}$ contains individual records whose influence should be removed while preserving utility on retained data.  
In kernel QML, this often reduces to low-rank modifications of the Gram matrix $K_{jk}=\kappa_\phi(x_j,x_k)$ and efficient Sherman–Morrison–Woodbury updates; shallow, aligned kernels tend to dominate generalization and admit fast row/column re-estimation~\cite{wang2025qkar}.  
In variational models, edits localize to layers of $U(\theta)=\prod_\ell U_\ell(\theta_\ell)$ guided by sensitivity; a first-order approximation
\begin{equation}
\theta^{\setminus \mathcal{S}} \approx \theta - H^{-1}\nabla_\theta \mathcal{L}_{\mathcal{S}}(\theta),
\qquad
H=\nabla^2_\theta \mathcal{L}_{\mathcal{D}}(\theta),
\label{eq:if-scope}
\end{equation}
captures the local effect of $\mathcal{S}$, with $H^{-1}$ replaced in practice by damped or block-diagonal QFI preconditioners for stability on NISQ hardware~\cite{daka2025nv}.

At the \emph{class level}, we remove geometry associated with one or more labels and retain the rest.  
Mixed-state generative classifiers enable component-level deletion without retraining entire models~\cite{useche2025qgc}.  
Search over circuit families with tunable expressivity supports targeted suppression of class-specific features~\cite{beaudoin2025qfusion}.  
Compact QVC ensembles further allow pruning of subcircuits and light retraining of combiner layers for sequential tasks~\cite{moll2025cyber}.

At the \emph{client level}, relevant to QFU, forgetting combines parameter edits with state-theoretic maps.  
A local CPTP channel acts on the client register, followed by partial retraining of the global model; distributed QLSTM and kernelized pipelines implement such workflows with secure coordination~\cite{chen2025dqlstm,hsu2025qklstm,mantha2025pilot}.  
Applications in security and materials show that lightweight hybrids with small latent spaces permit rapid reinitialization and refitting~\cite{sridevi2025hqc,frehner2025qae-ts,kang2025mtv}.  
Choosing scope is thus a design decision: sample-level for precise removal~\cite{wang2025qkar}, class-level for generative or structural shifts~\cite{useche2025qgc}, and client-level when privacy or provenance dominates~\cite{chen2025dqlstm,mantha2025pilot}.  
Auditability and secure aggregation constraints should inform this choice~\cite{moreau2025qai,aizpurua2025vqa}.

\subsection{Guarantees: Empirical, Certified, and Differential Privacy}
\label{subsec:guarantees}

Guarantees make forgetting credible.  
\emph{Empirical} guarantees rely on reproducible trends across seeds, backends, and datasets.  
QSVM and kernel hybrids show stable accuracy on compressed medical datasets such as CompressedMediQ and QProteoML~\cite{chen2025compressedmediq,priyadharshini2025qproteoml}, and robustness persists under decoherence and backend variability~\cite{mahdian2025qsvment,khalil2025pegasosqsvm}.  
These studies validate observed forgetting behavior, although they do not bound worst cases.

\emph{Certified} guarantees quantify drift to a retrained counterfactual.  
For kernels, a rank-$r$ change $K\mapsto K-\Delta$ yields prediction bounds,
\begin{equation}
|f^{\setminus \mathcal{S}}(x)-f(x)| \le
\|k(x)\|\cdot\|(K+\lambda I)^{-1}\Delta(I+U)^{-1}\alpha\|_2,
\label{eq:kernel-cert}
\end{equation}
strengthened by precompression~\cite{chen2025compressedmediq}.  
For variational models,
\begin{equation}
\|\theta^{\setminus\mathcal{S}}-\theta\|_2
\le
\|H^{-1}\|_2\,\|\nabla_\theta \mathcal{L}_{\mathcal{S}}(\theta)\|_2,
\label{eq:if-cert-2}
\end{equation}
and Lipschitz readouts convert parameter gaps into output deviations.  
Quantum amplitude estimation adds additive risk certificates with query complexity $\widetilde{\mathcal{O}}(1/\varepsilon\log(1/\delta))$~\cite{delejarza2025qint}.  
Hybrid QEKLR decouples quantum features from convex readouts, enabling post-hoc checks via margins and ROC curves~\cite{misra2025qeklr}.

\emph{Differential privacy (DP)} complements these by bounding leakage under post-processing.  
For PQCs, clip parameter-shift gradients, add Gaussian noise at the optimizer or aggregator, and track composition with a moments accountant.  
Kernel pipelines privatize rows and aggregated statistics.  
At the client level, clipping plus secure aggregation provides federated privacy~\cite{mahdian2025qsvment,khalil2025pegasosqsvm}.  
Explainability and QAE-derived uncertainty support DP audits, while Shapley analyses confirm removal at block and gate granularity~\cite{delejarza2025qint,priyadharshini2025qproteoml,heese2025shapley}.

\subsection{Mechanisms: Reset, Fisher Guidance, and Gradient Reversal}
\label{subsec:mechanisms}

Mechanisms enact the edit.  
\emph{Reset + partial retraining} reinitializes high-influence subspaces and fine-tunes on $D_s$.  
QcNet-like hybrids reset quantum front-ends while retaining classical back-ends~\cite{gunasekar2025qcnet}; regression-style QCL models exploit few-shot re-optimization~\cite{mohamed2025qcl}.  
These strategies are shallow and NISQ-friendly.

\emph{Fisher-guided} methods prioritize sensitive coordinates.  
The QFIM,
\begin{equation}
F_{ij}=\mathrm{Re}\!\left[\langle\partial_i\psi|\partial_j\psi\rangle
-\langle\partial_i\psi|\psi\rangle\langle\psi|\partial_j\psi\rangle\right],
\label{eq:qfim-tax}
\end{equation}
identifies parameters that carry most influence.  
A diagonal or block-diagonal natural step,
\begin{equation}
\theta_i \leftarrow \theta_i - \eta\, F_{ii}^{-1}\,\nabla_{\theta_i}\mathcal{L}_{\mathcal{S}}(\theta),
\label{eq:fisher-tax}
\end{equation}
reduces global distortion while achieving local removal.  
Empirically, Fisher-aware hybrids outperform random resets and pair well with boosting-style reweighting (AdaBoost.Q)~\cite{chen2025adaboostq}.

\emph{Gradient reversal and inversion} target correlations directly.  
Parameterized quantum combs approximate inverse processes and minimize Bures divergence to unwind unwanted effects~\cite{mo2025pqcomb}.  
Ensembles combine reversal with Fisher rebalancing to preserve accuracy under component erasure~\cite{chen2025adaboostq}.  
Libraries such as \texttt{sQUlearn} package QFIM access, reversal hooks, and reset utilities for practical deployments~\cite{kreplin2025squlearn}.

\begin{table}[t]
\centering
\caption{Unlearning mechanisms in QML/QFL and their practical trade-offs.}
\label{tab:mechanisms}
\begin{tabular}{lllll}
\toprule
Mechanism & Operation & Certifiability & NISQ fit & DP link \\
\midrule
Influence+QFI & Eq.\ \eqref{eq:qfi-unlearn} & High (local) & Medium & Clipping, noise \\
Reset+Partial & Block reinit + fine-tune & Medium & High & Low sensitivity \\
Grad.\ reversal & Adversarial negation & Low & Experimental & Depends on clip \\
Reservoir readout & Linear forgetting & Medium & High & Statistic DP \\
Kernel alignment & Map change + MMD & Empirical & High & Gram DP \\
\bottomrule
\end{tabular}
\end{table}

Together, these techniques constitute the core operational toolkit for quantum unlearning.
They differ in their theoretical guarantees, compatibility with noisy intermediate-scale quantum (NISQ) devices, and interaction with differential-privacy (DP) controls.
Influence- and QFI-weighted methods offer the most interpretable local guarantees through geometry-aware updates,
while reset-based strategies provide high NISQ feasibility with limited retraining.
Adversarial or gradient-reversal schemes remain exploratory but reveal how inversion dynamics can be repurposed for forgetting.
Reservoir and kernel approaches trade analytical rigour for scalability and ease of implementation, relying on statistical DP mechanisms or kernel perturbations for privacy.
Table~\ref{tab:mechanisms} summarizes these methods and their practical trade-offs across certifiability, hardware compatibility, and DP alignment.

\subsection{System and Hardware Contexts}
\label{subsec:system-hardware-contexts}

Systems determine orchestration; hardware sets feasibility.  
Pure QML maximizes coherence yet faces trainability limits.  
Rotation-equivariant ans\"atze curb barren plateaus~\cite{sein2025rotation}; QGANs achieve strong fidelity with hardware-aware scheduling and mitigation~\cite{pajuhanfard2025qgan}; and quantum neural states support block-local edits~\cite{zhang2025nqs}.  
In QFU, modular circuits and block partitioning enable selective reset or reversal with classical controllers managing proofs and audits~\cite{sein2025rotation,zhang2025nqs}.  
Temporal hybrids like QK-LSTM embed sequences in quantum feature spaces and converge reliably across rounds~\cite{hsu2025qklstm}.  
Topology-aware orchestration reduces wall-clock unlearning time without sacrificing fidelity~\cite{phalak2025qualiti}.

Hardware choices shape the menu of mechanisms.  
Simulators offer reproducibility and controlled ablations; QRDR provides logarithmic scaling for compression~\cite{yang2025qrdr}.  
Superconducting QPUs deliver fast parameterized gates and low-latency resets~\cite{ding2025pqg}.  
Trapped ions sustain deep coherence for high-fidelity inversion and noise-precision calibration~\cite{bu2025qsp}.  
Annealers implement unlearning as Hamiltonian reconfiguration by flipping local fields~\cite{seong2025hamiltonian}.  
Deep variational circuits with curvature-aware optimizers support structured forgetting in vision workloads~\cite{xu2025vqnn}.  
These options define a practical design space in which guarantees and mechanisms must be matched to device realities.

With this scaffold in place, Sec.~\ref{sec:from-learning-to-unlearning} translates the axes into concrete algorithms: influence- and QFI-weighted updates, reset-and-finetune schedules, kernel alignment, and process-level inversion, together with diagnostics and certification tools that tie forgetting to counterfactual retraining.

\section{Journey From Machine Learning to Machine Unlearning}
\label{sec:from-learning-to-unlearning}

\subsection{Influence- and QFI-weighted unlearning}
\label{subsec:influence-unlearning}

The goal is simple to state and hard to achieve: remove the effect of $D_r$ while preserving utility on $D_s=D\setminus D_r$.  
Classically, the first-order impact of deleting $x_i$ at an optimum $\theta^\star$ is well captured by the influence-function step
$\Delta\theta_i \approx -H_{\theta^\star}^{-1}\nabla_\theta L(\theta^\star;x_i)$~\cite{koh2017understanding}.  
This view supports certified corrections that bound the gap to a counterfactual retrain~\cite{guo2020certified,bourtoule2021machine}.  
In the quantum setting, PQCs expose exact hardware gradients via the parameter-shift rule~\cite{schuld2019evaluating}, so per-sample or per-client gradients are measurable.  
Replacing $H^{-1}$ with the inverse QFI aligns the step with the local geometry and improves stability on NISQ devices~\cite{stokes2020quantum}:
\begin{equation}
\theta \leftarrow \theta - \eta\,F(\theta)^{-1}
\bigg(\frac{1}{|B|}\sum_{x\in B}\nabla_\theta\mathcal{L}(\theta;x)\bigg),
\label{eq:qfi-unlearn}
\end{equation}
for mini-batches $B\subseteq D_r$.  
Sensitivity governs effectiveness.  
Local QFI spectra and effective dimension shape generalization~\cite{khanal2025qfi-bounds}.  
Expressivity and gate choices modulate gradient scales and plateau risk~\cite{liu2025expressibility}.  
Layerwise growth preserves gradient signal~\cite{qi2025dlvqnn}, while dissipative VQAs mix reset and stochastic gates to steer toward low-sensitivity regions~\cite{ilin2025dvqa}.  
Beyond variational learning, reservoirs move adaptation to a linear readout where marked samples can be down-weighted~\cite{beaulieu2025qrc}.  
Training “rewinding’’ can undo influential segments using Fisher analysis~\cite{demiranda2025rewinding}.  
Graph-partitioned control exposes modular subcircuits for local edits~\cite{liu2026mpgpqoc}.  
QCNN encoders stabilize gradients so that forgetting can target the encoder~\cite{tang2025qcnn-enc}.  
Hybrid encoders amplify structure in small data and enable edits in embedding space~\cite{jin2025proton}.  
In federated settings, these steps compose with masked sharing and topology-aware coordination~\cite{bhatia2024bhi}.

Operationally, we judge success by contraction to the retrained reference.  
Let $\theta^{\setminus D_r}$ be the optimum on $D\setminus D_r$.  
Effective unlearning reduces
\begin{equation}
\mathcal{D}\!\big(\rho(\theta),\rho(\theta^{\setminus D_r})\big),
\qquad
\mathcal{D}\in\{\text{trace distance},\ 1-\text{fidelity}\},
\label{eq:dist-obj}
\end{equation}
because these metrics reflect observable proximity.  
CPTP monotonicity ensures device noise and partial measurements further contract $\mathcal{D}$, yielding conservative, hardware-grounded certificates on NISQ devices.  
This link will underpin the evaluation metrics in Sec.~\ref{subsec:metrics}.

\begin{algorithm}[t]
\caption{QFI-weighted influence unlearning (QMU-I)}
\label{alg:qmu-i}
\begin{algorithmic}[1]
\REQUIRE Trained parameters $\theta$, forget set $D_r$, retained set $D_s$, step $\eta$, clipping $C$
\STATE Estimate local QFI $F(\theta)$ (e.g., parameter-shift Fisher blocks)
\FOR{mini-batches $B\subseteq D_r$}
  \STATE $g \leftarrow \frac{1}{|B|}\sum_{x\in B}\nabla_\theta \mathcal{L}(\theta;x)$ \hfill (parameter shift)
  \STATE $g \leftarrow \mathrm{clip}(g, C)$ \hfill (stability, DP sensitivity)
  \STATE $\Delta \leftarrow F(\theta)^{-1} g$ \hfill (geometry-aware preconditioning)
  \STATE $\theta \leftarrow \theta - \eta\,\Delta$ \hfill (update)
  \STATE \textbf{optional:} trust region projection with $\|\Delta\|_{F(\theta)} \le \tau$
\ENDFOR
\STATE Fine-tune on $D_s$ with natural gradient and early stopping
\STATE Audit $\mathcal{D}\big(\rho(\theta),\rho(\theta^{\setminus D_r})\big)$ with probes and membership-risk tests
\end{algorithmic}
\end{algorithm}

Among the candidate strategies, QMU-I (Algorithm~\ref{alg:qmu-i}) offers the most interpretable and verifiable path for geometry-aware forgetting. It provides a balance between shot efficiency and formal trace-distance certifiability, making it particularly suitable for NISQ and federated settings.

\subsection{Parameter reinitialization and partial retraining}
\label{subsec:reinit-partial}

Reinitialization localizes change while protecting the useful hypothesis.  
Partition $\theta=(\theta_{\rm keep},\theta_{\rm forget})$, reset the latter from a reference $p_0$, and fine-tune a small set $\Theta(\mathcal{S})$ on $D_s$ with a stabilizer $R$ that preserves observables or kernels.  
Where to reset depends on where information lives.  
Quantum kernels suggest most influence sits in the feature map; changing maps can outperform expensive kernel retraining~\cite{alvarez2025kernel}.  
Memetic and architecture search reveal sensitive groups~\cite{ardila2025memoqcd}.  
Quantum pointwise convolutions provide natural reset points for cross-channel mixing~\cite{ning2025qpointwise}.  
Hybrid QNNs often match classical accuracy with far fewer parameters, so quantum-layer resets yield large effect with small retraining~\cite{bischof2025entity}.  
Model families admit tailored edits: QRBMs and generators reset priors/latents while keeping critics~\cite{sinno2025qrbm,ghosh2025guardians}; graph and molecular pipelines reset encoders while preserving heads~\cite{an2025tensorenc,lu2025qegnns}; hybrid quantum–neural wavefunctions reset quantum blocks and use a classical corrector~\cite{li2025hybridwf}.

A geometric view clarifies why this works.  
With QFI $F(\theta)$, a local surrogate for the counterfactual gap is
$\delta(\theta)\approx \|\theta-\theta^{\setminus D_r}\|_{F(\theta)}^2$.  
Resetting coordinates with large QFI eigenvalues maximizes forgetting per parameter.  
Natural-gradient fine-tuning within $\Theta(\mathcal{S})$ then recovers utility without leaving the unlearning manifold.  
We monitor trace distance or infidelity; for kernels, MMD and alignment provide efficient surrogates~\cite{alvarez2025kernel}.  
For generators, provenance signals from hardware noise help attest that the post-unlearning model is not the pre-unlearning instance, and privacy audits probe membership and inversion risk~\cite{ghosh2025guardians}.  
These diagnostics connect directly to the reporting checklist in Sec.~\ref{subsec:metrics}.

\subsection{Empirical and certified forgetting}
\label{subsec:empirical-vs-certified}

Evidence comes in two forms.  
\emph{Empirical} tests probe membership inference on $D_r$, accuracy on held-out $D_r$, and retention on $D_s$.  
Large QML studies show mixed trends and strong dependence on feature maps~\cite{mohammadisavadkoohi2025survey,qi2025iscas}.  
Noise can aid forgetting through contractivity~\cite{nguyen2020robust}.  
Even compact Lorentz-equivariant QGNNs, while accurate, need removal tests under shift~\cite{jahnin2025lorentz}.

\emph{Certified} forgetting targets the counterfactual.  
If $\theta'$ is the post-unlearning model and $\bar\theta$ the retrained optimum, a certificate ensures
\begin{equation}
\mathcal{D}\big(\rho(\theta'),\rho(\bar\theta)\big)\le \varepsilon.
\label{eq:cert-gap}
\end{equation}
Classical tools use stability, low-rank corrections~\cite{guo2020certified}, SISA partitioning~\cite{bourtoule2021machine}, or deletion-aware sampling~\cite{ginart2019making}.  
In QML, measurement design tightens generalization~\cite{li2025measurement}; QFI spectra and effective dimension supply geometry-aware bounds that complement \eqref{eq:qfi-unlearn}.  
Kernel surrogates yield empirical certificates: if $K'$ aligns with $\bar K$ while suppressing $D_r$-specific structure, loss correlates with the alignment gap~\cite{alvarez2025kernel}.  
Hybrid QEKLR separates feature maps from convex readouts for clean post-hoc checks~\cite{misra2025qeklr}.  
Explainability strengthens audits: Shapley values reveal block importance~\cite{heese2025shapley}; drops in targeted blocks support removal claims.  
Reverse engineering of transpiled circuits remains a risk~\cite{ghosh2025re}, so cryptographic provenance is valuable.  
In federated settings, resilience matters.  
Auction-based selection and ring topologies harden aggregation~\cite{lee2025auction}.  
Certified statements can hold at the aggregate when client updates meet Lipschitz-like conditions, which aligns with QFL evidence~\cite{qi2025iscas,mohammadisavadkoohi2025survey}.  
These certification practices foreshadow the formal and ethical agenda in Sec.~\ref{sec:open-challenges}.

\subsection{QFL architectures and their unlearning levers}
\label{subsec:qfl-architectures}

Architecture dictates leakage channels and the cost of edits.  
In the \emph{model plane}, kernel QFL shares feature maps and aggregates readouts, often preferable to heavy retraining~\cite{alvarez2025kernel}.  
Near-optimal kernel PCA scales dimensionality reduction~\cite{wang2025qkpca}.  
In oncology, lower-expressivity encodings prove more robust on hardware~\cite{repetto2025breast}.  
Variational QFL trains PQCs or hybrids: QAEs yield compact features~\cite{asaoka2025qae}; QCNN hybrids improve NISQ feature learning~\cite{li2025cqhcnn}; topology-first search finds circuits before parameter refinement~\cite{su2025tdqas}; and QAOA depth motivates depth-aware designs~\cite{chicano2025qaoa}.  
Reservoir and generative QFL reduce on-hardware backprop and exploit noise shaping~\cite{ahmed2025qrc,tian2025qgan}.

Topology and aggregation complete the picture.  
Clients update $\theta_i$ locally and an aggregator forms $\theta^{t+1}$ from $\{\theta_i^{t+1}\}$.  
Rings reduce central memory and match emerging quantum networks~\cite{wang2024quantum}; chain-based QFL removes the server~\cite{gurung2025ccqfl}; slimmable QFL adapts width and measurement to channels~\cite{park2025slimqfl}; CC-QFL uses shadow tomography with classical clients~\cite{song2024ccqfl}; FedQLSTM accelerates temporal tasks~\cite{chehimi2024fedqlstm}.  
Privacy levers include gradient hiding over quantum channels~\cite{li2024privacy} and PQFL with homomorphic encryption and QDP~\cite{mashetty2025pqfl}.  
Non-sequential hybrids such as QPIE transfer classical knowledge to flatten Fisher spectra~\cite{guo2025qpie}; provenance signals aid audits in generative settings~\cite{ghosh2025guardians}.  
Architectures that concentrate influence into identifiable modules make unlearning cheaper: feature maps and encoders for kernels and PQCs, readouts for reservoirs, and priors for QGAN/QRBM.  
These levers connect directly to the scope and mechanism choices of Sec.~\ref{sec:taxonomy} and to privacy accounting in Sec.~\ref{subsec:secure-agg-dp}.

\subsection{Client- and entanglement-level forgetting}
\label{subsec:client-entanglement}

Client removal should avoid full retraining.  
Let $\rho(\theta)$ be the aggregate state on $\mathcal{H}=\bigotimes_i\mathcal{H}_i$ and $\bar\rho=\rho(\theta^{\setminus c})$ the counterfactual without $D_c$.  
A local LOCC/CPTP map
\begin{equation}
\mathcal{U}_c(\rho)=\big(\mathcal{E}_c\otimes \mathsf{Id}_{\neg c}\big)(\rho),
\qquad
\mathcal{E}_c\approx \mathrm{Tr}_c\circ \mathcal{D}_c,
\label{eq:client-map}
\end{equation}
neutralizes client coherences, and contractivity gives
\begin{equation}
\mathcal{D}\!\big(\mathcal{U}_c(\rho),\bar\rho\big)\le \mathcal{D}\!\big(\rho,\bar\rho\big).
\label{eq:dp}
\end{equation}
Entanglement-aware levers support this step.  
Encoder resets reinitialize feature maps and measurements with light server-side tuning.  
Feature-space mapping links expressivity to encoding and entanglement patterns~\cite{walia2025featuremap,hangun2025wind}.  
MERA-like pruning collapses correlations~\cite{zhang2025mera}.  
Neuromorphic perceptrons thin entanglement and mitigate plateaus~\cite{bravo2025qperceptron}.  
Concrete handles exist across families: drop client Gram blocks and re-embed with lower-expressivity encodings in kernels~\cite{repetto2025breast}; localize influence in encoders for targeted resets in PQC hybrids~\cite{asaoka2025qae,li2025cqhcnn}; forget at the readout in reservoirs~\cite{ahmed2025qrc}; remove latent modes in qGANs with adaptive training~\cite{tian2025qgan}; and harden outputs with DP before cryptographic use~\cite{ahn2025qryptgen}.  
Audits should report membership risk for $D_c$, alignment shifts, entanglement reduction, and observable norms linked to generalization~\cite{hagelueken2025excited}.  
Qudits and mixed registers alter crosstalk and twirling costs, so client-local subspaces should be chosen with tracing efficiency in mind~\cite{venturelli2025qudit}.  
These practices foreshadow fairness and governance issues in Sec.~\ref{sec:open-challenges}.

\subsection{Secure aggregation and differential privacy}
\label{subsec:secure-agg-dp}

Security defines what can leak and under which coalitions.  
Secure aggregation computes $G=\sum_i g_i$ without revealing individuals, while DP bounds what any released mechanism $\mathcal{M}(G)$ reveals:
\begin{equation}
\Pr[\mathcal{M}(D)\!\in\!S]\le e^{\varepsilon}\Pr[\mathcal{M}(D')\!\in\!S]+\delta,
\label{eq:dp-def}
\end{equation}
and quantum post-processing preserves DP.  
For QFL, options include homomorphic encryption with quantum updates (PQFL)~\cite{mashetty2025pqfl}, quantum gradient hiding with low communication~\cite{li2024privacy}, and ring/chain topologies that avoid central aggregation~\cite{wang2024quantum,gurung2025ccqfl}.  
Middleware can align circuit cutting with secure rounds~\cite{mantha2025pilot}.  
For PQCs, clip parameter-shift gradients and add Gaussian noise with $\sigma\!\propto\! C\sqrt{2\log(1.25/\delta)}/\varepsilon$; measurement design reduces sensitivity and tightens bounds~\cite{li2025measurement}.  
For kernels, fix maps, clip clients, and privatize Gram aggregation~\cite{alvarez2025kernel}.  
Mixed-state classifiers support DP over sufficient statistics~\cite{useche2025qgc}.  
Time-series QFL composes DP across long horizons with a moments accountant~\cite{chen2025dqlstm,hsu2025qklstm,frehner2025qae-ts}.  
Generative flows and qGANs combine clipping with privatized discriminator statistics and remain robust to adaptive injections~\cite{laird2025gflownet,tian2025qgan}.  
Explainable stacks assist audits~\cite{don2025qrlxai}, and security analytics confirm feasibility on real QPUs with modest parameter counts~\cite{sridevi2025hqc}.

A practical round is straightforward: clip, securely aggregate (ring masks or HE), add Gaussian noise, update the moments accountant, and schedule influence-based resets (Sec.~\ref{subsec:reinit-partial}).  
Dynamic aggregation recovers utility at the same privacy budget~\cite{qu2025daqfl}.  
Domain pipelines can embed these steps into scientific circuits~\cite{kang2025mtv}.  
We will report these privacy and audit settings alongside accuracy and physics-aware distances in Sec.~\ref{subsec:metrics}.

\subsection{Complexity, feasibility, and failure modes}
\label{subsec:complexity-failure}

Feasibility hinges on shot cost and communication.  
With $n$ qubits, depth $d$, $p$ parameters, and $S$ shots, a parameter-shift gradient costs $O(S)$ per parameter and $O(pS)$ per batch.  
Block-diagonal QFI adds a constant factor; dense QFI scales as $O(p^2)$ unless sparsity is exploited.  
Thus natural steps with Fisher blocks fit NISQ by sharing shifts and amortizing $F(\theta)$.  
In QFL with $m$ clients, per-round cost is $O(p)$ for PQCs or $O(N_k)$ for kernel statistics; slimmable sharing, sketching, and ring aggregation reduce load.  
Influence+QFI (Alg.~\ref{alg:qmu-i}) works best for shallow circuits and few observables.  
Reset+partial retraining is resilient when gradients plateau.  
Kernel and reservoir routes minimize shots but require careful map or readout design.

Typical failure modes are known.  
Barren plateaus call for layerwise growth, encoder resets, or dissipative gates~\cite{qi2025dlvqnn,ilin2025dvqa}.  
Mis-targeted resets benefit from attributions and kernel alignment~\cite{alvarez2025kernel,heese2025shapley}.  
Gradient and kernel leakage is mitigated by clipping, secure aggregation, and DP accounting~\cite{li2024privacy,mashetty2025pqfl}.  
Utility collapse on $D_s$ is checked by trust-region natural gradients and early stopping.  
Diagnostics should report membership risk on $D_r$ vs.\ $D_s$, probe-set distances in \eqref{eq:dist-obj}, kernel MMD/alignment gaps, entanglement monotones across $(c,\neg c)$, and attribution drops in targeted blocks~\cite{alvarez2025kernel,heese2025shapley}.  
QFL audits add per-round $\varepsilon$ and secure-aggregation alarms.  
These diagnostics and costs will be standardized in Sec.~\ref{subsec:metrics} and motivate the open challenges in Sec.~\ref{sec:open-challenges}.

\section{Evaluation Metrics and Datasets}\label{sec:eval}

\subsection{Metrics and Reporting}\label{subsec:metrics}

Evaluating quantum unlearning requires both task-level and physics-aware metrics that jointly measure utility preservation, effective forgetting, and system reliability.

\begin{itemize}[leftmargin=*]
    \item \textbf{Operational distances.} Model forgetting is quantified through operational metrics that capture state distinguishability. The trace distance $D_{\mathrm{tr}}(\rho,\sigma)=\tfrac{1}{2}\|\rho-\sigma\|_1$ and infidelity $1-F(\rho,\sigma)$ evaluate proximity between the post-unlearning model and the counterfactual retrained model.
        \item \textbf{Utility and robustness.} Task-level metrics ensure that forgetting does not degrade retained performance. Common measures include accuracy, AUC, and calibration on the retained dataset $D_s$, as well as robustness under adversarial poisoning or distribution shift (before and after unlearning). For kernel-based models, Maximum Mean Discrepancy (MMD) and alignment gaps provide efficient proxies for distributional drift.
        \item \textbf{Capacity and sensitivity.} Quantum Fisher information (QFI) spectra and effective dimension quantify model sensitivity to perturbations, guiding geometry-aware unlearning. Measurement Frobenius norms correlate with tighter generalization bounds~\cite{li2025measurement}. Feature attributions such as Shapley values further offer auditable explanations of which circuit blocks carry forgettable information~\cite{heese2025shapley}.
        \item \textbf{Privacy.} Differential privacy (DP) metrics formalize confidentiality in distributed or federated unlearning. Report $(\varepsilon,\delta)$ parameters using a moments accountant, together with clipping norms, noise multipliers, and secure aggregation settings, enabling reproducible privacy guarantees across clients.
        \item \textbf{Reproducibility.} For physical credibility and replicability, report experimental details such as:
            \begin{itemize}[leftmargin=*]
              \item random seeds, backend type, and device calibration;
              \item number of qubits, circuit depth, and shots per measurement;
              \item transpilation strategies, gate error rates, and queue latencies.
            \end{itemize}
            Such documentation enables hardware-to-hardware comparability and ensures traceable evaluation pipelines.

        \item \textbf{Temporal behavior.} It is an essential diagnostic for assessing whether forgetting stabilizes over time or continues to drift.  The forgetting curve $F(t)$ characterizes the evolution of model distinguishability as unlearning proceeds.  Ideally, $F(t)$ decays toward a stable equilibrium $\rho_f$ that approximates the counterfactual retrained state.  
\end{itemize}

\subsection{Benchmark Datasets}
\label{subsec:benchmarks}

Benchmarking quantum unlearning requires diversity across domains and circuit complexities.  
We group representative datasets into three thematic categories.

\begin{itemize}[leftmargin=*]
\item \textbf{Vision and imaging.} Compressed neuroimaging datasets such as \textit{CompressedMediQ} combine CNN or PCA front-ends with quantum SVM or kernel back-ends~\cite{chen2025compressedmediq}. Complex imaging datasets drive VQ-DNN and variational architectures optimized via curvature-aware updates~\cite{xu2025vqnn}.

\item \textbf{Physics and quantum tasks.} Quantum SVM-based entanglement detection benchmarks hardware fidelity across backends~\cite{mahdian2025qsvment}. Hamiltonian clustering tasks on annealers offer realistic settings for hybrid optimization and unlearning validation~\cite{seong2025hamiltonian}.

\item \textbf{Health, proteomics, and text.} Minority-cohort proteomics (QProteoML) test fairness under differential forgetting~\cite{priyadharshini2025qproteoml}.Fake-news classification using \textit{PegasosQSVM} includes simulation-to-hardware discrepancy analysis~\cite{khalil2025pegasosqsvm}.  
Temporal health pipelines benchmark recurrent quantum models such as QK-LSTM and QLSTM~\cite{hsu2025qklstm,chen2025dqlstm}.
\end{itemize}

\begin{table}[t]
\centering
\caption{Benchmark datasets and domains for evaluating quantum unlearning across complexity and hardware depth.}
\label{tab:benchmarks}
\begin{tabular}{p{0.26\columnwidth}p{0.30\columnwidth}p{0.20\columnwidth}p{0.24\columnwidth}}
\toprule
\textbf{Domain} & \textbf{Dataset / Task} & \textbf{Qubits / Depth} & \textbf{Purpose} \\
\midrule
Vision & MNIST, CIFAR-10 & 6–12 / 20–40 & Pattern forgetting, visual drift \\
Astronomy & HTRU-1 / HTRU-2 pulsar data & 4–8 / 15–25 & Outlier removal under QK-SVM \\
Energy & Battery-health degradation & 8–10 / 30–60 & Temporal QLSTM unlearning \\
Chemistry & QSPR, QM9 molecular descriptors & 6–12 / 35–45 & Property re-tuning and privacy deletion \\
Healthcare & QProteoML, EEG / fMRI & 10–16 / 25–50 & Patient-level and class-level forgetting \\
Text & Fake-news (PegasosQSVM) & 4–6 / 15–20 & Semantic forgetting in embeddings \\
\bottomrule
\end{tabular}
\end{table}

These benchmarks collectively provide cross-domain coverage of structured, temporal, and physical data, ensuring that forgetting performance can be measured both algorithmically and experimentally.

Table~\ref{tab:benchmarks} summarizes representative domains, typical circuit sizes, and the role each dataset plays in assessing QMU.  
This mix enables controlled ablations (e.g., scope and mechanism choices from Sec.~\ref{sec:taxonomy}), end-to-end trials with federated orchestration (Sec.~\ref{sec:from-learning-to-unlearning}), and physics-aware audits that support the open challenges in Sec.~\ref{sec:open-challenges}.  
The result is a reporting template that ties utility, privacy, and distinguishability to concrete workloads and devices, closing the loop between theory and practice.

\section{Open Challenges and Research Agenda}
\label{sec:open-challenges}

Quantum Machine Unlearning (QMU) sits at the intersection of information theory, computation, and governance.  
The foundations in Sec.~\ref{sec:foundations} show that forgetting must respect CPTP dynamics and contract operational distances.  
The taxonomy in Sec.~\ref{sec:taxonomy} organizes concrete levers across scope, guarantees, and mechanisms.  
What remains is to convert these ingredients into proofs, scalable systems, interpretable outcomes, and auditable practice.  

\subsection{Formal Proofs of Quantum Forgetting}
\label{subsec:formal-proofs-quantum-forgetting}

Classical guarantees rely on retraining equivalence, influence corrections, or stability bounds.  
Quantum settings require a different contract: suppression of \emph{retrievable} information within CPTP maps.  
A useful target is to certify that fidelity and trace distance cross pre-set margins between pre- and post-unlearning states, that is $F(\rho,\rho')\le \epsilon$ and $D_{\mathrm{tr}}(\rho,\rho')\ge \delta$, for task-specific thresholds~\cite{nhlapo2025zerosvm,klymenko2025architecturalpatterns,lin2025veriqr}.  
These conditions pair naturally with the data-processing inequality and turn hardware noise into conservative evidence rather than a nuisance.

Parameterized quantum circuits offer structure for proofs.  
Modular architectures allow selective resets or stochastic reinitializations of \emph{forgetting blocks} that isolate the contribution of marked data or clients~\cite{nhlapo2025zerosvm}.  
After a reset, vanishing local gradients $\partial_{\theta_i}\mathcal{L}=0$ signal orthogonality to forgotten modes and support geometric certificates.  
Architectural patterns strengthen this path.  
Decoherence-mediated forgetting uses controlled channels $\mathcal{E}_p$ to raise entropy on targeted subspaces.  
Variational reset reinitializes blocks and bounds post-tuning drift.  
Hybrid modularity separates quantum edits from classical feedback to prevent re-entanglement~\cite{klymenko2025architecturalpatterns}.  
Bridging practice and proof will also need empirical-to-formal pipelines.  
Hybrid models that couple classical encoders with variational circuits retain accuracy after local resets and supply testbeds for certificate design~\cite{guha2025resq}.  
Verification tools such as \textsc{VeriQR}, designed for robustness, can reason about reachability before and after unlearning and check disjointness of state sets within a contractive metric~\cite{lin2025veriqr}.  
These steps align theory with thermodynamics and move QMU toward machine-checkable guarantees.

\subsection{Scalability and Secure Architectures}
\label{subsec:scalability-secure-architectures}

Scalability is a joint hardware–systems problem.  
NISQ limits on depth and connectivity require modular subsystems with classical orchestration and explicit verifiability.  
Reservoir computing offers a principled scaffold.  
Classical and quantum reservoirs are universal for fading-memory functionals, so local readout resets can approximate global retraining with bounded error~\cite{monzani2025universality}.  
This supports distributed unlearning without full model restart.

Fully quantum generative models are another lever.  
QGANQB keeps both generator and discriminator quantum, enabling native process-level unlearning and avoiding classical bottlenecks~\cite{ding2025qganqb}.  
Quantum imaging adds a complementary knob.  
Compact encodings such as FRQI and NEQR permit localized spectral edits (Fourier or Haar) that implement class- or region-specific forgetting with polynomial resources~\cite{farooq2025quantumimage}.  
Security must be baked into these designs.  
Controlled entanglement interfaces and modular middleware isolate subsystems and limit cross-node leakage in federated settings~\cite{klymenko2025architecturalpatterns}.  
This architectural security complements QDP and homomorphic transport and supports the secure-composition guarantees needed for QFU at scale.

\subsection{Post-Unlearning Interpretability}
\label{subsec:post-unlearning-interpretability}

Erasure should not obscure why the model still works.  
Interpretability after unlearning is an information allocation problem: reduce $I(\rho;\rho')$ while preserving explanatory power under relevant observables.  
Self-adaptive quantum kernel PCA can re-optimize components to suppress variance tied to removed samples and keep embeddings faithful to the retained subspace~\cite{wang2025saqkpca}.  
Physics-informed hybrids, such as TE-QPINNs, embed constraints from governing equations so that deletion does not violate domain knowledge~\cite{berger2025teqpin}.  
Architectural lineage also helps.  
Evolutionary records in hybrids like HQCNN-REGA trace which blocks were altered and support audit-ready explanations~\cite{mukhanbet2025hqcnnrega}.  
Formally, one can target $\min I(\rho;\rho')$ subject to $\mathcal{I}(\rho',\mathcal{O})\ge \tau$ for chosen observables $\mathcal{O}$ and threshold $\tau$.  
This framing matches the sensitivity tools in Sec.~\ref{sec:from-learning-to-unlearning} and yields interpretable, deletion-aware reports.

\subsection{Ethics and Societal Implications}
\label{subsec:ethics-societal-implications}

QMU must earn trust through verification and fairness.  
Verifiable delegation allows clients to check remote computations and supports auditable unlearning on untrusted hardware~\cite{yang2025verifiablevqa}.  
Biomedical deployments underscore the need for human oversight and reproducibility when QSVCs approach classical accuracy~\cite{passo2025ibmbrain}.  
Hardware fairness is a new axis.  
Noise varies by device, which can induce unequal reliability.  
Calibration transparency, cross-device reproducibility, and bounds on trace-distance divergence should therefore be part of governance checklists~\cite{wang2025noisykernel}.  
In federated settings, deletion proofs and DP accounting must be first-class artifacts, not afterthoughts.  
These practices connect directly to the metrics in Sec.~\ref{subsec:metrics} and prepare the ecosystem for standardized audits.

\begin{figure}[t]
\centering
\includegraphics[width=0.8\columnwidth]{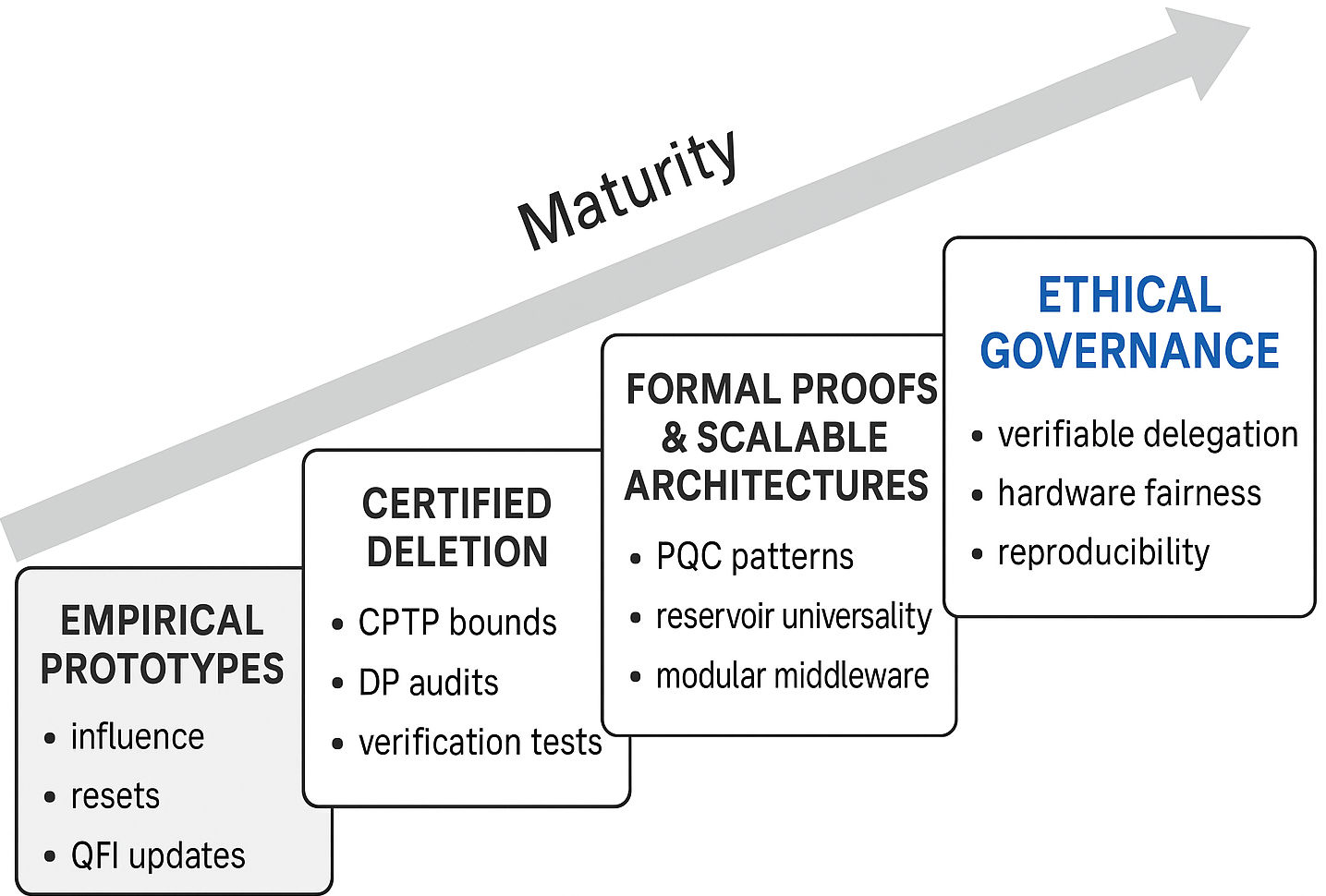}
\caption{A roadmap from empirical unlearning to certified, scalable, and ethically governed QMU.  
Arrows indicate dependencies between proofs, systems, interpretability, and governance.}
\label{fig:roadmap}
\end{figure}

Table~\ref{tab:future} summarizes the agenda and links each objective to enablers and measurable outcomes.  
Figure~\ref{fig:roadmap} places these steps on a trajectory from empirical heuristics to certified, scalable, and ethically governed systems.  
The next section distills these directions into actionable recommendations and ties them back to the taxonomy, mechanisms, and datasets presented earlier.

\begin{table}[t]
\centering
\caption{Future research directions in Quantum Machine Unlearning (QMU): objectives, enablers, and measurable outcomes.}
\label{tab:future}
\begin{tabular}{p{0.23\columnwidth}p{0.43\columnwidth}p{0.28\columnwidth}}
\toprule
\textbf{Objective} & \textbf{Methodological Enablers} & \textbf{Expected Outcomes / Metrics} \\
\midrule
Provable forgetting &
Modular PQCs; CPTP-based certificates; \textsc{VeriQR}-style reachability~\cite{nhlapo2025zerosvm,klymenko2025architecturalpatterns,lin2025veriqr} &
Certified bounds on $F$ and $D_{\mathrm{tr}}$; machine-checkable proofs. \\[1.2ex]
Scalable QML/QFU &
Reservoir universality; fully quantum GANs; secure modular middleware~\cite{monzani2025universality,ding2025qganqb,klymenko2025architecturalpatterns} &
Sub-linear retraining cost; elastic depth/width; efficient client-level deletion. \\[1.2ex]
Post-unlearning interpretability &
SAQK-PCA; TE-QPINNs; evolutionary lineage (HQCNN-REGA)~\cite{wang2025saqkpca,berger2025teqpin,mukhanbet2025hqcnnrega} &
Observable-preserving edits; quantitative interpretability indices; audit trails. \\[1.2ex]
Ethical governance &
Verifiable delegation; calibration fairness; reproducible diagnostics~\cite{yang2025verifiablevqa,wang2025noisykernel,passo2025ibmbrain} &
Standardized reports; backend-agnostic audits; accountable QFU pipelines. \\
\bottomrule
\end{tabular}
\end{table}

\section{Conclusion}
\label{sec:conclusion}

Quantum machine unlearning (QMU) represents a foundational step toward making privacy and accountability intrinsic features of quantum intelligence rather than external requirements. As learning migrates to quantum substrates, the principles of data deletion must conform to the physical constraints of quantum mechanics, where information cannot be destroyed but only redistributed through completely positive trace-preserving (CPTP) channels. This paper unifies the theoretical and practical dimensions of this challenge, framing forgetting as the contraction of distinguishability between quantum models trained with and without specific data. It integrates insights from quantum information theory, thermodynamics, and privacy-preserving computation, demonstrating how mechanisms such as QFI-weighted influence updates, parameter reinitialization, reservoir resets, and kernel alignment can achieve localized forgetting with verifiable stability under noisy intermediate-scale quantum (NISQ) conditions. On a systems level, we connect these algorithmic foundations to secure architectures, embedding Quantum Differential Privacy (QDP), Quantum Key Distribution (QKD), and homomorphic encryption within federated frameworks that support client-level deletion and entanglement disentanglement. The proposed taxonomy and evaluation protocols establish measurable standards for empirical and certified unlearning, ensuring both physical feasibility and ethical traceability. Despite these advances, open challenges persist: formal proofs of quantum forgetting, scalable architectures for heterogeneous hardware, interpretable post-unlearning models, and auditable governance mechanisms that extend accountability across quantum networks. The convergence of these research directions will transform the right to erasure from a regulatory mandate into a verifiable property of physical computation. In this vision, QMU emerges not only as a technical innovation but as a philosophical redefinition of intelligence itself, where learning and forgetting coexist as two sides of the same information-theoretic process, guaranteeing that the preservation of privacy and the pursuit of knowledge remain fundamentally compatible in the quantum era.

\section*{Acknowledgments}

\bibliography{sn-bibliography}


\end{document}